\documentclass[universe,review,accept,moreauthors,pdftex]{Definitions/mdpi}
\usepackage{color}

\firstpage{1}
\pubvolume{1}
\issuenum{1}
\articlenumber{0}
\pubyear{2021}
\copyrightyear{2020}
\datereceived{}
\dateaccepted{}
\datepublished{}
\hreflink{https://doi.org/} % If needed use \linebreak
\Title{Magnetospheric physics of magnetars}

\TitleCitation{Magnetospheric physics of magnetars}

\def\apj{ApJ}
\def\apjl{ApJL}
\def\apjs{ApJ Suppl. Ser.}
\def\apss{Astroph.Sp.Sci.}
\def\aap{A\&A}

\def\aapr{A\&AR}
\def\araa{ARAA}
\def\mnras{MNRAS}

\def\natas{Nat. As.}
\def\pasj{PASJ}

\def\raa{RAA}

\Author{Hao Tong\orcidA{}}

\AuthorNames{Hao Tong}

\AuthorCitation{Tong, H.}
\address{School of Physics and Materials Science, Guangzhou University, Guangzhou 510006, China}

\corres{Correspondence: tonghao@gzhu.edu.cn}

\abstract{Several aspects of the magnetospheric physics of magnetars are summarized, including: GeV and hard X-ray emissions of magnetars, timing behaviors during magnetar outburst (soft X-ray observations), optical/IR observations of magnetars, radio emission of magnetars, and accreting magnetars. A unified picture for pulsars and magnetars are adopted, especially wind braking of magnetars, magnetar+ fallback disk systems, twisted dipole magnetic field, and accreting low magnetic field magnetars etc. It is pointed out that magnetars are related to a broad range of astrophysical phenomena.}


\begin{document}
%%%%%%%%%%%%%%%%%%%%%%%%%%%%%%%%%%%%%%%%%%
%\setcounter{section}{-1} %% Remove this when starting to work on the template.

%main materials of the three sections are put here
\section{Introduction}

Magnetars may be young and strongly magnetized neutron stars. In the first place, magnetars are pulsars. Observationally, there is a continuous distribution from normal pulsars to magnetars. Therefore, every model for magnetars should return back to the normal pulsar case in some parameter space. Otherwise, it can not be the truth. Secondly, magnetars are a special kind of pulsars. Their energy output may be dominated by the magnetic energy release. In this respect, many of the physics of solar magnetic field can be applied to the case of magnetars, as already been done by in previous works.  This forms our basic understanding of magnetars. 

Pulsars are rotating magnetized neutron stars \cite{Lyne2012}. They are good clocks in the universe. The basic magnetospheric physics of pulsars try to answer ``why they are good clocks?" This question can be divided into two parts: (1) the radiation mechanism of pulsar's multi-wave emissions, from radio to GeV etc, (2) what's the spin-down mechanism of pulsars\footnote{For accreting pulsars and accreting magnetars, the question is their spin-up mechanism}? These two questions can also be applied to the study of magnetars. The position of magnetars on the pulsar $P$-$\dot{P}$  diagram can be found in Figure 1. Magnetars have longer period and larger period-derivative than that of normal pulsars. The characteristic age $\tau= P/(2\dot{P})$ and characteristic magnetic field $B_c = 3.2\times 10^{19} \sqrt{P \dot{P}}$ give a crude estimation of the source's age and magnetic field. Constant lines of characteristic age and characteristic magnetic field is plot in Figure 1. A general picture of magnetars can be got from it: magnetars are young (age $10^4-10^5 \ \rm yr$) and strongly magnetized (magnetic field $\ge 10^{14} \ \rm G$) neutron stars, compared with that of normal pulsars. 

\begin{figure}
  \centering
  \includegraphics[width=0.5\textwidth]{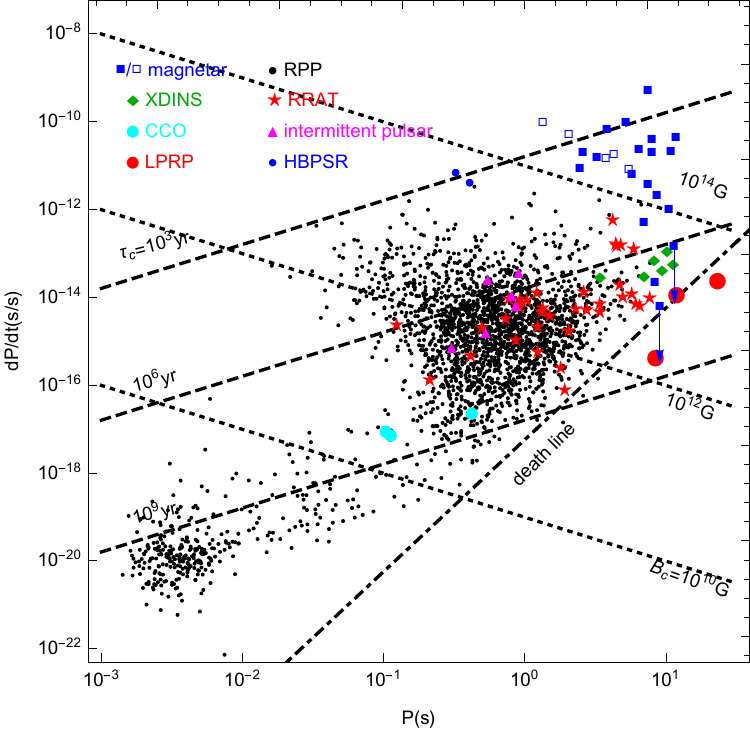}
  \caption{Magnetars on the $P-\dot{P}$ diagram. Blue squares are magnetars, empty blue squares are radio emitting magnetars. Updated from Figure 1 in \cite{TongHuang2020}, see the caption there for the meaning of various pulsar-like objects.}\label{fig_PPdot}
\end{figure}

{\it A brief history of magnetars.}  Before the discovery of magnetars, normal rotation-powered pulsars had already been discovered in 1967 \cite{Hewish1968}, and accretion-powered X-ray pulsars in binary systems had already been discovered in 1971 \cite{Giacconi1971}. The history of magnetar research can be dated back to 1979 with the discovery of a giant flare in SGR 0526-66 in LMC \cite{Mazets1979}. However, its nature and physics was uncertain at that time.  The anomalous X-ray pulsar 1E 2259+586 was discovered in 1981 \cite{Fahlman1981}. Again, its nature was unclear at that time. The idea of strongly magnetized neutron star and the acronym ``magnetar" was proposed in 1992 \cite{Duncan1992,Usov1992,Paczynski1992}. After the timing of SGR 1806-20 \cite{Kouveliotou1998} and giant flare from SGR 1900+14 \cite{Hurley1999} in 1998, the idea of magnetars became more and more popular. After the discovery of radio, optical/IR and persistent hard X-ray emission from magnetars around 2006 \cite{Camilo2006,Wang2006,Gotz2006}, the study of magnetars entered into the multiwave era (previously, the observations of magnetar were mainly in the soft X-ray band and hard X-ray only for the bursts).  Later on, the the research of magnetars becomes more and more diverse, e.g., discovery of low magnetic field magnetars in 2010 \cite{Rea2010}, accreting magnetars in ultra-luminous X-ray pulsars in 2014 \cite{Bachetti2014}, fast radio burst from magnetars in 2020 \cite{Chime2020,Bochenek2020,Lin2020}. 

\subsection{Overview of magnetars}

In the early references, magnetars are grouped into two subclasses: anomalous X-ray pulsars (dubbed as AXPs), and soft gamma-ray repeaters (dubbed as SGRs). Anomalous X-ray pulsars are ``anomalous" because their X-ray luminosity are higher than their rotational energy loss rate and they show no binary signature. Therefore, their energy budget is a puzzle at that time (at a time when only rotation-powered pulsars and accretion-powered pulsars are known). Soft gamma-ray repeaters got their name in comparison with typical gamma-ray bursts: the typical photon energy of soft-gamma repeaters is lower than that of gamma-ray bursts, and they can show recurrent bursts (i.e., not one-off events). Today (2023), both anomalous X-ray pulsars and soft-gamma repeaters are believed to be the same class observationally. They are believed to be magnetars, i.e. neutron stars powered by their magnetic energy release. In the following, we will mainly use the name ``magnetar". 

{\it Quantum critical field.} In magnetar researches, the quantum critical magnetic field is often employed. It is defined as the magnetic field when the electron cyclotron energy equals its rest mass energy: $B_q = m_e^2 c^3/(e \hbar) = 4.4 \times 10^{13} \ \rm G$. The meaning of quantum critical field is that when dealing with microscope physics (e.g., conduction coefficient) in such strong field quantum electrodynamics (dubbed as QED) should be employed. The non-relativistic Schr\"{o}dinger equation is no longer valid in  such strong magnetic field. The macroscopic physics is unchanged, e.g. the magnetohydrodynamics is still valid in the neutron star crust. 

{\it Traditional picture of magnetars.} A traditional picture of magnetars can be found in earlier reviews of magnetars (e.g., \cite{Mereghetti2008}). Based on multiwave observations at that time, a magnetar is thought to be: (1) a young neutron star (since its is in association with supernova remnants and massive star clusters), (2) with strong dipole magnetic field, e.g., higher than the quantum critical value $B_{\rm dip} > B_q =4.4\times 10^{13} \ \rm G$. The dipole field is mainly responsible for the braking of magnetars. (3) with even stronger multipole field, e.g., $B_{\rm mul} = 10^{14}-10^{15} \ \rm G$. It is the multipole field which is responsible for magnetar's multiwave emissions, e.g., burst, super-Eddington luminosity and persistent emissions. This traditional picture is very nice. But it may be too simple to represent the complicated and diverse  observations of magnetars. 

A wonderful example of magnetar observations is their giant flares (only three up to present, Fig. 7 in \cite{Mereghetti2008}). A giant flare of magnetars consists of (1) a spike and a pulsating tail (the tails can last for hundreds of seconds), (2) $10^4$ times super-Eddington luminosity can be reached during the tail (i.e., $10^{42}  \ \rm erg \ s^{-1}$). A magnetic field as high as $10^{15} \ \rm G$ may serve as the energy power of the giant flare and may cause the super-Eddington luminosity during the tail phase \cite{Paczynski1992}. In our opinion, this is the strongest evidence for the existence of magnetars (e.g, compared with that of fallback disks). Magnetars are restless compared with that of normal rotation-powered pulsars. The magnetar SGR 1806-20 not only has a giant flare, but also has many bursts \cite{Woods2007,Younes2017}. During these period, the frequency derivative (i.e., spin-down torque) can vary by a factor of several, even order of magnitude. Therefore, their position on the $P$-$\dot{P}$ diagram (Figure 1) changes with time, which is uncommon for normal pulsars. Magnetars are not just neutron stars with strong magnetic fields. Magnetars are special because they can show various activities and variabilities. 

The discovery of low magnetic field magnetars clearly shows that the tradition picture is incomplete \cite{Rea2010,Rea2012,Zhou2014}. Magnetars not only have bursts (including giant flares), but also can have outbursts: increase of persistent X-ray flux by two or three orders, then decay in the following months or years \cite{Alford2016}. During the outburst, magnetars can have variable spectra and timing behaviors, transient radio emissions (\cite{Camilo2006}, the first transient magnetar and the first radio emitting magnetar XTE J1810-197). Up to now, many magnetars show outbursts and their light curve are generally an exponential decaying function or some more complicated form \cite{CotiZelati2018}. However, if looking into the details, every magnetar has its own peculiarity. For example, the magnetar SGR 1935+2154 is not such outstanding by looking at its outbursts. However, it showed Galactic fast radio bursts, accompanied by X-ray burst \cite{Chime2020,Bochenek2020,Lin2020,Li2021}. 

{\it Magnetospheric physics of magnetars.} The reason why magnetars can have many activities while normal pulsars can't may be that: magnetars have a twisted magnetosphere compared with that of normal pulsars \cite{Thompson2002}. The magnetic field of normal pulsars may be mainly dipole, with no twist and may be considered as the ground state of the magnetic field configuration. While the magnetic field of magnetars may be a twisted dipole, plus some local twisted multipole field. A twisted magnetic field has free energy compared with a dipole field. The release of this magnetic free energy may responsible for the burst and multiwave emissions of magnetars \cite{Beloborodov2009,Tong2019,YuHuang2013}. 

This review mainly lists several observational aspects of magnetars and points out the possible magnetospheric physics behind them. It is in no way complete. More can be found in \cite{Mereghetti2008}(still intriguing at present), \cite{Mereghetti2015,Turolla2015,Kaspi2017,Pons2019,Esposito2021,Igoshev2021} etc. A general overview of magnetars has been given in this section. The following sections will provide more details on different aspects and can be read/skipped by relevance. 

\section{Multiwave emissions of magnetars}

\subsection{GeV emission of magnetars}

By applying the outer gap model to magnetars, it is expected that some of them should have gamma-ray emission and may be detected by Fermi \cite{Cheng2001}. However, Fermi/LAT observation of 4U 0142+61 (the brightest magnetar in X-rays) resulted in non-detection \cite{SasmazMus2010}. Fermi/LAT observation of all magnetars also reported non-detection \cite{Abdo2010}. By applying the outer gap model with updated input to magnetars \cite{Tong2010,Tong2011}, it is found that three of the magnetars should have been seen by Fermi. And the Fermi upper limits of 4U 0142+61 is already below the theoretical spectral energy distributions. The conflicts between the outer gap in the case of magnetars and Fermi observations is confirmed by later more observations \cite{Li2017}. 

The Fermi observations of magnetars imply that either (1) the outer gap model is wrong (which is unlikely considering the observation of gamma-ray pulsars \cite{Smith2023}), or (2) the traditional magnetar picture is wrong. It can not be excluded that AXPs/SGRs are actually accreting systems \cite{Chatterjee2000,Alpar2001,Ertan2009,TongXu2011}. Another solution is that: magnetars may be wind braking \cite{Tong2013}. In the wind braking scenario, magnetars do not have a strong dipole field, and the vacuum gaps (e.g., outer gap) can not exist in the case of magnetars. 

There are also very high energy (e.g., TeV) observations of magnetars \cite{Aleksic2013,Abdalla2021}.  The general upper limit is about $1\%$ Crab unit. Considering that the Crab rotational energy loss rate is about $10^{38} \ \rm erg \ s^{-1}$, the typical X-ray luminosity (which is powered by the magnetic energy release and much higher than their rotational energy loss rate) of magnetars is about $10^{35} \ \rm erg \ s^{-1}$ which is $1/1000$ of the Crab rotational energy loss rate. Therefore, a $1\%$ Crab unit upper limit is not constraining. In the future, GeV and TeV emission of magnetars during bursts (including giant flares)  may be one thing that can be expected. 

\subsection{Hard X-ray emission of magnetars}

Magnetars can have persistent hard X-ray emissions \cite{Gotz2006}. This is rather unexpected, considering the power law component of magnetar's soft X-ray component is rather steep. In contrast, the hard X-ray emission of magnetars have a rather flat spectra. Their total luminosity is comparable with that of soft X-rays. Therefore, the hard X-ray emission forms a distinct component in addition to the soft X-ray component (the soft X-ray component is mainly composed of a blackbody+power law). Combined with Fermi non-detection in the GeV range ($>100 \ \rm MeV$, \cite{Abdo2010,Enoto2011}), the spectra of magnetars is expected to have a cut-off around $\sim 1 \ \rm MeV$. The exact cut-off energy is unknown at present. 

There are many theoretical modelings of magnetar's hard X-ray emissions (\cite{Wang2014} and references therein). Possible candidates includes: bremsstrahlung, resonant Compton scattering, synchrotron process, or bulk motion Comptonization in the accretion model etc. Both the magnetospheric models and bulk motion Comptonization model are possible \cite{Trumper2010,Guo2015}. Insight-HXMT observations may further constrain the cut-off and spectra of magnetar's hard X-ray emissions \cite{Wang2014}. 

\subsection{Timing behaviors during outburst (soft-X-ray observations)}

The soft-X-ray band is one of the two main channels that we learn about magnetars (the other is radio observations). During the soft X-ray outburst, magnetars show various kinds of timing and spectra variabilities. At the same time, transient hard X-ray emission (for \cite{Enoto2021} for a recent example) and transient radio emission may also be detected during outburst. We will focus on magnetar's timing behaviors during outburst. 

{\it How magnetars are spun-down down?}. There are many timing events in magnetars, including: varying period derivative, low magnetic field magnetars (or low period-derivative magnetars), anti-glitches, negative correlations between X-ray luminosity and period derivative (period derivative reflects the spin-down torque of the magnetar).  For example, repeated and delay torque variations are seen several times in magnetar 1E 1048.1$-$5937 \cite{Archibald2015,Archibald2020}, varying spin-down rate is also found using radio observations of the magnetar PSR J1622$-$4950 \cite{Levin2012}, a possible decreasing spin-down rate is also reported in the low magnetic field magnetar Swift J1822.3$-$1606 \cite{Scholz2014}. This raises the question: ``How magnetars are spun-down?". A physical model for this question should answer: (1) why there are so many timing  events in and mainly in magnetars?  (2) How to unify the spin-down mechanism of pulsars and magnetars? 

{\it Various modelings for magnetars.} For the spin-down mechanism and related physics of magnetars, there  are various modelings, employing or not employing the magnetar model. These includes \cite{TongXu2014}: (1) neutron+ twisted magnetosphere model \cite{Thompson2002,Beloborodov2009,Tong2019}, (2) Wind braking model of magnetars \cite{Tong2013}, (3) Coupled magnetic and thermal evolution \cite{Vigano2013} (the first three modelings are in the magnetar domain), (4) fallback disk model \cite{Chatterjee2000,Alpar2001,Ertan2009}, (5) Accretion induced star-quake model \cite{Xu2006}, (6) Quark nova remnant \cite{Ouyed2007}, (7) white dwarf model for AXPs and SGRs \cite{Paczynski1990,Malheiro2012} etc. These modelings may share some common merits. There are also various subsequent developments for every model.  

The wind braking model of magnetars focus on the timing behaviors of magnetars. The general picture of wind braking is similar for  both normal pulsars and magnetars: (1) the particle in the magnetosphere will experiences acceleration and subsequent radiation. This results in the star's multiwave emissions. (2) When flowing out, this particle component will also take away the rotational energy of the neutron star. This results in the spin-down of the pulsar. The same particles will contribute both to the radiation and spin-down of the neutron star. Therefore, a correlation between the emission and timing bahaviors is naturally expected in the wind braking model. In the case of normal pulsars, the spin-down is made up by the sum of dipole radiation and particle component (\cite{XuQiao2001,KouTong2015} and references therein). There are various winds in the universe: solar wind, stellar wind (Wolf-Rayet stars which will result in Ib, Ic supernova, high-mass X-ray binaries which are neutron stars accreting the wind of its binary companion). The particle (and electromagnetic field) outflow in the case of pulsars and magnetars is named as ``wind". The existence of pulsar wind nebulae clearly demonstrates the wind of pulsars \cite{Gaensler2006}. 

{\it Wind braking of magnetars.} In the wind braking model of magnetars \cite{Tong2013},  the rotational energy loss rate is enhanced by the particle wind $\dot{E}_w = \dot{E}_d (L_p/\dot{E}_d)^{1/2}$, where $\dot{E}_w$ and $\dot{E}_d$ are rotational energy loss rate due to particle wind and magnetic dipole radiation, respectively, $L_p$ is the particle wind luminosity. The particle wind is dominated by magnetic energy release in the case of magnetars. It may be comparable with the X-ray luminosity, which can be significant larger than the magnetar's rotational energy loss rate:  $L_p \sim L_x \gg \dot{E}_{\rm rot}$. In this case, the required dipole magnetic field will be much smaller than the magnetic dipole braking assumption (i.e., the characteristic magnetic field). In the case of normal pulsars, the particle wind is due to rotational energy loss rate $L_p \sim \dot{E}_{\rm rot}$. Then the rotational energy loss due to particle wind is comparable with the magnetic dipole radiation. Therefore, wind braking of magnetars can unify the spin-down mechanism of normal pulsars and magnetars.  The wind braking model of magnetars had also be proposed earlier \cite{Harding1999}. However, when Harding et al saw that a strong dipole magnetic field is not needed in the wind braking model, they said ``the magnetar model must be abandoned" as the penalty of the wind braking model. However, one point they did not realize is that there are two kinds of magnetic field in magnetars (dipole field and multipole field).  In the wind braking model, a magnetar is a neutron star with strong multipole field, with or without a strong dipole field (which does not play a significant role). Once this point is realized, many challenging observations of magnetars can be understood \cite{Tong2013}. It also has clear predictions: a magnetism-powered pulsar wind nebula and a braking index smaller than three. 

Several subsequent observations of magnetars can also be explained in the wind braking model. For example: (1) the timing behavior of the low magnetic field magnetar \cite{Rea2010,TongXu2012}. (2) Possible decreasing period derivative (i.e. spin-down torque) of the low magnetic field magnetar Swift J1822.3$-$1606 \cite{Scholz2014,TongXu2013}. (3) Negative correlation between the X-ray luminosity and spin-down torque of the Galactic center magnetar \cite{Kaspi2014,Tong2015}. (4) The anti-glitch in the magnetar 1E 2259+586 can also be understood naturally in the wind braking scenario \cite{Archibald2013,Tong2014}. (5) Possible magnetar wind nebula around the magnetar Swift J1834.9-0846 \cite{Younes2016,Tong2016}. Glitches of normal pulsars have been studied for over 50 years. In the case of magnetars, anti-glitches are discovered. This again demonstrates the peculiarity of magnetars. Magnetars can provide us many things unexpected in normal pulsars (FRBs, anti-glitch, fallback disks etc). 

The reason why magnetars can have so many timing and radiative events may be that their magnetic field is a twisted dipole field instead of simple dipole field. The idea of twisted magnetic field has been discussed earlier in the case of the solar magnetic field \cite{Wolfson1995}. Later it has been applied to the case of magnetars \cite{Thompson2002,Beloborodov2009}. From a geometrical point of view, the twisted can be westward or eastward.  More importantly is that a twisted magnetic field carries magnetic free energy. The long term flux decay of transient magnetars may due to untwisted of a globally twisted magnetic field \cite{Tong2019}. In a globally twisted magnetic field, the magnetar may have large polar caps. This will have many sequences for the radio and X-ray emissions of magnetars. The calculations in both local twist or globally twist is rather complicated. A simplified model (toy model) is developed for the flux decay, shrink hot spot and delayed spin-down torque of magnetar outburst \cite{TongHuang2020}. A toy model is easy to use, especially for observers.

\subsection{Optical/IR observations of magnetars: fallback disks}

Some of the ejected material of a supernova may fallback and form a disk around the central compact star. This is the idea of fallback disks. In the case of pulsars, the propose of a fallback disk has a long history\cite{Michel1988}. However, no fallback disk is found observationally. Optical/IR observations of the magnetar 4U 0142+61 revealed possible existence of a fallback disk \cite{Wang2006}. The idea of finding a fallback disk is achieved in a magnetar. Both the magnetar model and the fallback disk model claim their success for the fallback disk around 4U 0142+61.  

{\it Magnetar+fallback disk system.} The combination of magnetar+fallback disk made a success in explaining the central compact object (dubbed as CCO, cyan cycles in Figure 1) inside supernova remnant RCW 103. This compact has a pulsation period about $6.6$ hours \cite{DeLuca2006}. It is confirmed to be a magnetar (by it magnetar-like burst and outburst, \cite{D'Ai2016,Rea2016}). This will make the magnetar inside RCW 103 a very special magnetar. Compared with other magnetars, other central compact objects, normal pulsars, and accreting neutron stars, the magnetar inside RCW 103 has the longest pulsation period at that time. A combination of magnetar+fallback disk may explain its long pulsation period \cite{Tong2016b}. A high disk mass ($\sim 10^{-5} \ \rm M_{\odot}$) and high dipole field ($\sim 5\times 10^{15} \ \rm G$) is required to explain a period about $2\times 10^{4} \ \rm s$.  Later discovery of long period radio pulsars (dubbed as LPRPs, red cycles in Figure 1, where sources with period longer than 23.5s are not shown. See Figure 1 in \cite{Tong2023b} for updates.) may also be magnetar+fallback disk systems \cite{HurleyWalker2022,HurleyWalker2023,Tong2023a,Tong2023b}. Another possibility for long period radio pulsar is that they are white dwarfs \cite{Katz2022}. It is interesting to note that, the white dwarf model is also an alternative to the magnetars model for AXPs and SGRs (The fallback disk model is also originally proposed to beat the magnetar model). The fallback disk may be relevant to many other aspects of pulsars and magnetars (nulling, braking index, precession etc, \cite{Tong2020} and references therein). 

\subsection{Radio emission of magnetars}

Radio observations found most of the pulsars up to now \cite{Han2021}. Radio emission of magnetars also reveal a wealth of information about their physics and provide links between magnetars and normal pulsars.  Among the about 30 magnetars up to now \cite{Olausen2014}, 6 magnetars are observed to have radio emissions (chronological order, emphasizing the radio polarization aspect):(1) XTE J1810$-$197 (first transient magnetar and first radio emitting magnetar \cite{Camilo2006,Camilo2007,Kramer2007}, revived later \cite{Dai2019}), (2) 1E 1547.0$-$5408 \cite{Camilo2008}, (3) PSR J1622$-$4959 (\cite{Levin2012}, revived later \cite{Camilo2018}), (4) SGR 1745$-$2900 (Galactic centre  magnetar \cite{Eatough2013,Yan2015}), (5) Swift J1818.0$-$1607 (may be a transition object \cite{Champion2020,Lower2021,Huang2021}), (6) SGR 1935+2154 (emitting Galactic FRBs \cite{Zhu2023}). The radio emitting high magnetic field pulsar (blue dots in Figure 1) PSR J1119$-$6127 is also reported to have magnetar activities \cite{Dai2018}. These observations of magnetar radio emissions provides clues and links to the physics of FRBs \cite{Fonseca2020,Luo2020}. 

The first radio emitting magnetar XTE J1810$-$197 and later more sources tell us that \cite{Camilo2006,Dai2019}: (1) magnetar radio emissions have a flat spectra, (2) they are high variable (flux, pulse profile, timing etc), (3) they are transients (disappear during the outburst decaying phase, and may revive during the next outburst). The radio-loud magnetar PSR J1622$-$4950 shows a decreasing polarization position angle with time \cite{Levin2012}, which implies a timing evolving magnetosphere of magnetars. Swift J1818.0$-$1607 has a steep spectra at first and flat spectra later, which may be a transition object between normal pulsars and magnetars \cite{Huang2021}. The bright and narrow single pulse and flat polarization angle is similar to other magnetars and that of FRBs \cite{Champion2020,Lower2021,Fonseca2020}. The polarization position angle also changes slope which has never been observed in normal pulsars, which again require a dynamic magnetosphere of magnetars \cite{Lower2021}. A dynamic polarization position angle is also found in FRBs \cite{Luo2020}, which may imply similar physics are happening there. 

In summary, the radio emissions of magnetars are highly variable (flux, pulse profile, timing, polarization, and position angle etc). The position angle depends on the magnetic field geometry of the neutron star. A timing varying position angle may indicate a timing varying magnetic field in magnetars. This is consistent with the untwisting picture of magnetar outburst. The question is: how to model the position angle for a complex field geometry? Once obtained, this model can be applied to both magnetars and FRBs. 

{\it Rotating vector model for magnetars.} Assuming a globally twisted dipole magnetic field, the magnetic field geometry can be approximated analytically \cite{Tong2019}. Employing spherical geometry or differential geometry, the modification of the rotating vector model (a model for the position angle) in the case of magnetars can be obtained analytically \cite{Tong2021}. Once another magnetic field geometry is obtained, given the toroidal field, the modification to the position angle can also be approximated. Therefore, every magnetospheric model for magnetars should calculate its field geometry and compare with the radio observations of magnetars. In the presence of multipole field, the appearance and disappearance of multipole  field may cause a changing slope of the position angle (\cite{Lower2021}).

\subsection{Accreting magnetars}

Magnetars are just a special kind of pulsars. Since there are accreting neutron stars in binary systems\cite{Giacconi1971}, then it is natural to expect that there are magnetars in binary systems. The question is: how can we identify possible signatures of magnetars in a binary system? From the X-ray spectra perspective, magnetars have a flat hard X-ray spectra, compared with rotation-powered pulsars and accretion-powered pulsars \cite{Mereghetti2015}. From the physical point of view (as discussed above), the key difference between magnetars and rotation-powered pulsars is magnetar's multipole field. It is not their position on the $P-\dot{P}$ diagram (i.e., not the surface dipole field). Therefore, we must find evidences for strong multipole field in accreting systems, in order to say that they are accreting magnetar candidates. Possible evidences for strong multipole field include \cite{TongWang2014}: (1) magnetar burst, (2) hard X-ray tail etc. One thing that is rather unexpected is the discovery of ultra-luminous X-ray pulsars, which may be super-Eddington accreting magnetars in binary systems \cite{Bachetti2014}. 

Ultraluminous X-ray sources are super-Eddington (for a stellar mass object) point sources offset from the galactic centre. Previously, the ultra-luminous X-ray sources are thought to be intermediate mass black holes or super-Eddington accreting stellar mass black holes \cite{King2023}. The detection of pulsation (a pulsation period modulated by the orbital motion) from an ultra-luminous X-ray source confirms their neutron star nature. Like that of normal accreting neutron stars, the ultra-luminous X-ray pulsar is also observed to be spinning up \cite{Bachetti2014}. Then the problem (or difficulty) of ultra-luminous X-ray source is two fold: how to explain their super-Eddington luminosity (e.g., $10^{40} \ \rm erg \ s^{-1}$) and their spin-up rate (i.e., even if we do no know the super-Eddington mechanism, such a huge accretion flow will result in a very large spin-up torque, which is much larger than the observed value)? The answer is accreting magnetars. It is proposed in both the observational paper \cite{Bachetti2014} and subsequent modelings \cite{Tong2015b,Shao2015,Pan2016}.
%(Eksi et al. 2014; Lyutikov 2014; Kluzniak \& Lasota et al. 2014; Tong 2014; Chrostodoullou et al. 2014; Dall'Osso et al. 2014; Mushtukov et al. 2015; Pan et al. 2015; King \& Lasota 2016; Fragos et al. 2015; Shao \& Li et al. 2015). 

{\it Accreting low magnetic field magnetar.} In the accreting low magnetic field magnetar scenario, the super-Eddington luminosity is due to the presence of strong multipole field (e.g., $10^{14} \ \rm G$). The rotational behaviors of the ultra-luminous X-ray pulsar is due to interaction of its much lower dipole field (e.g., $10^{12} \ \rm G$) with the accretion flow \cite{Tong2015b}. The idea of accreting low magnetic field magnetar is consistent with the studies of isolated magnetars (aged magnetars are more likely to be  low magnetic field magnetars). The propose of accreting low magnetic field is consistent with later observations \cite{Israel2017} and theoretical models \cite{TongWang2019,Chen2021}. Accreting magnetars may be related to the formation of some peculiar millisecond pulsars \cite{Pan2016}. 

{\it Three kinds of accreting magnetars.} At present, the slowest pulsation X-ray pulsar is AX J1910.7+0917 with a period about 10 hours \cite{Sidoli2017}. The magnetar inside RCW 103 (with a period of 6.6 hours) is the now the second slowest rotating neutron star. Similar to the magnetar inside RCW 103, AX J1910.7+0917 may be an accreting magnetar with low mass accretion rate in a binary system. If ultra-luminous X-ray pulsars are super-Eddington accreting magnetars, then it is possible that there are also other accreting magnetar with lower  mass accretion rate. In our opinion \cite{TongWang2019}, there may be three kinds of accreting magnetars: (1) ultra-luminous X-ray pulsars may be accreting magnetars with high mass accretion rate. The high mass accretion rate may result in the decay of the magnetic field. Thus result in an accreting low magnetic field magnetars. (2) Slow pulsation X-ray pulsars (e.g., AX J1910.7+0917, 2S 0114+65, 4U 2206+54, super-giant fast X-ray transients) may be accreting magnetars with low mass accretion rate. (3) Slow pulsation X-ray pulsars in SMC (with periods 1000s) may be accreting magnetars with  intermediate  mass accretion rate. Considering that 4U 2206+54 is spinning down, while 2S 0114+65 is spinning up, it is possible that AX J1910.7+0917 is the linking source between 4U 2206+54 and 2S 0114+65. For the fallback disk accreting magnetar inside RCW 103, ultra-luminous X-ray pulsar, and slow pulsation X-ray pulsars, they may all be accreting magnetars. Accreting magnetars are also magnetars.

\section{Summary: magnetars in astrophysics}

At present, magnetars have a limited number of sources (about 30). Future more radio and X-ray observations may tell us more about magnetars \cite{Zhu2023,Taverna2022}. Magnetars are linked to a broad range of observations in astrophysics: (1) The typical examples are anomalous X-ray pulsars and soft-gamma ray repeaters. They may be magnetar candidates. (2) The X-ray dim isolated neutron stars  may be dead magnetars (blue diamonds in Figure 1). (3) For the central compact objects inside supernova remnants, they may be magnetar-in-waiting (i.e. anti-magnetar) or fallback disk accreting magnetars. (4) High magnetic field pulsars provide links between normal pulsars and magnetars (e.g., PSR J1846$-$0258 and PSR J1119$-$6127). (5) The existence of low magnetic field magnetars imply  that there may be magnetar  activities in normal pulsars in the future. (6) It is natural that there are also accreting magnetars in binary systems (e.g., LS I+61, superslow X-ray pulsars, ULX pulsars etc). (7) For the proposal of magnetars inside FRBs and GRBs, a definite period is required. Until then can we say that they are magnetars. 

\funding{This work is supported by National SKA Program of China (No. 2020SKA0120300) and NSFC (12133004).}

%\acknowledgments{ack...}
%{The author is grateful to ...... for helpful discussions.}

\conflictsofinterest{The author declare no conflict of interest.}

\appendixtitles{no} % Leave argument "no" if all appendix headings stay EMPTY (then no dot is printed after "Appendix A"). If the appendix sections contain a heading then change the argument to "yes".

\reftitle{References}

%\externalbibliography{yes}
%\bibliography{references}

\end{paracol}
\end{document}